\documentclass[reqno]{amsart}

\usepackage{amsthm,amsmath,amsfonts,amssymb,epsfig,graphicx,latexsym,float,color}
\usepackage{pgfplots}
\usepackage{tabularx}
\usetikzlibrary{plotmarks}
\usetikzlibrary{external}
\newtheorem{definition}{Definition}

\newtheorem{remark}[definition]{Remark}

\newtheorem{system}[definition]{System}

\begin{document}

\title[Modeling and Simulation of Curved Fibers in Dry Spinning Scenarios]{Modeling and Simulation of Curved Fibers in Dry Spinning Scenarios}

\author[Wieland et al.]{Manuel Wieland$^{1}$}
\author[]{Walter Arne$^{1}$}
\author[]{Nicole Marheineke$^{2}$}
\author[]{Raimund Wegener$^{1}$}

\date{\today\\
$^1$ Fraunhofer ITWM, Fraunhofer Platz 1, D-67663 Kaiserslautern, Germany\\
$^2$ Universit\"at Trier, Lehrstuhl Modellierung und Numerik, Universit\"atsring 15, D-54296 Trier, Germany}

\begin{abstract}
Dry spinning is characterized by the simultaneous production of multiple thin polymeric fibers in airflows and the evaporation of the solvent contained in these fibers. This involves fiber-air interactions as well as radial diffusion effects over the fiber cross-sections. Recent models for fibers in such production processes assume the fiber curves to be strictly uni-axial. But there are industrial setups showing airflow induced lateral movements of the fibers which creates the need for curved fiber models. In this paper we present a special Cosserat rod model that describes curved fibers in such dry spinning processes. Adapting the efficient simulation framework from the uni-axial fiber setting makes the simulation of industrial setups for multiple curved fibers with two-way coupled airflows possible, for which we present results for the first time in literature.
\end{abstract}
\maketitle
\noindent
{\sc Keywords.} dry spinning, fiber dynamics, parametric boundary value problem, homotopy method, heat and mass transfer, integral equations\\
{\sc AMS-Classification.} 34B08, 68U20, 35Q79, 76-XX
%
\setcounter{equation}{0} \setcounter{figure}{0} \setcounter{table}{0}
\section{Introduction}
Dry spinning is a production method for polymeric fibers. In dry spinning devices a polymer diluent solution is fed at constant and controllable rate through jet nozzles into a spinning duct, where the spun fibers are dried by a heated airflow. During this drying process solvent evaporates out of the jets and leads to thinning and solidification of the fibers.

Starting from uni-axial stationary one-dimensional dry spinning models introduced by \cite{ohzawa:p:1969, ohzawa:p:1970}, a two-dimensional model with concentration-dependent rheological laws covering the radial diffusion effects was employed in \cite{brazinsky:p:1975}. A comparative theoretical and experimental study of a cellulose acetate/acetone dry spinning system was presented in \cite{sano:p:2001}. The works of \cite{gou:p:2003, gou:p:2004} extended the models from viscous to viscoelastic material behavior. Concerning industrial devices up to several hundred fibers are spun simultaneously such that mutual fiber-air interaction effects have to be incorporated into the dry spinning models. In \cite{wieland:p:2019} we developed an efficient model simulation framework, which made a fully two-way coupled simulation of dry spun fibers immersed in an airflow feasible. Proceeding from a three-dimensional free boundary value problem for a viscous uni-axial radially symmetric stationary two-phase flow, we deduced a dimensionally reduced fiber model. We combined one-dimensional ordinary differential equations for fiber velocity and stress with two-dimensional advection-diffusion equations covering the cross-sectional variations of polymer mass fraction and fiber temperature. The efficiency of our proposed algorithmic procedure is based on the analytical solution of the advection-diffusion equations with Green's functions. These implicitly given solution expressions give rise to Volterra integral equations with singular integration kernel, which we solved efficiently by a product integration method. We particularly coupled them iteratively with a continuation-collocation algorithm for the ordinary differential equations.

Considering industrial dry spinning devices with lateral air inflow into the spinning duct, i.e., the airflow velocity is meanly perpendicular to the fiber direction, the assumption of straight fibers is no longer tenable due to expected lateral deflections. Therefore, we extend the model and simulation framework presented in \cite{wieland:p:2019} to the description of curved fibers in such processes. There are two classes of fiber models: string and more sophisticated Cosserat rod models. The rod models additionally contain angular momentum effects and the slenderness parameter $\varepsilon$. The limit $\varepsilon\rightarrow 0$ yields the corresponding string model. Since singularities restrict the solvability of viscous string models \cite{goetz:p:2008}, we focus on a fiber description with the help of viscous Cosserat rod models, which can be seen as a regularization of the corresponding string models \cite{arne:p:2011b}. Analogously to the uni-axial strings in \cite{wieland:p:2019}, we supplement the one-dimensional rod models for curved fibers with two-dimensional advection-diffusion equations describing the polymer mass fraction and temperature profiles of the fiber cross-sections and adopt the numerical procedure accordingly. In this paper we present simulation results for curved dry spun fibers with two-way coupled fiber-air interactions for the first time in literature.

This paper is structured as follows. In Sec.~\ref{sec:model} we present our dry spinning model for curved viscous fibers. The numerical solution framework is explained in Sec.~\ref{sec:numerics}. In Sec.~\ref{sec:ex} we show simulation results for an industrial dry spinning setup.

\section{Dry spinning model for curved fibers}\label{sec:model}
In the special Cosserat rod theory there are two constitutive elements: a curve ${\mathbf{r}}:\Omega\rightarrow\mathbb{E}^3$  specifying the jet position (e.g. midline) and an orthonormal director triad $\{{\mathbf{d_1}},{\mathbf{d_2}},{\mathbf{d_3}}\}:\Omega\rightarrow\mathbb{E}^3$ that is attached to the curve and characterizes the orientation of the cross-sections in the three-dimensional Euclidean space $\mathbb{E}^3$. Considering the stationary one-dimensional fiber domain $\Omega$ the fiber length $L$ is an a priori unknown parameter. It is common to represent the rod equations with respect to the director basis and an outer basis. This simplifies the discussion of geometric models and material laws. Besides the director basis  $\{{\mathbf{d_1}},{\mathbf{d_2}},{\mathbf{d_3}}\}$ we introduce the fixed (space- and time-independent) orthonormal outer basis $\{{\mathbf{a_1}},{\mathbf{a_2}},{\mathbf{a_3}}\} \subset \mathbb{E}^3$. The director and outer bases are related by the tensor-valued rotation ${\mathbf{R}}$, i.e.,\ ${\mathbf{R}}=\sum_{i=1}^3{\mathbf{a_i}}\otimes {\mathbf{d_i}}$. For any quantity ${\mathbf{y}}\in\mathbb{E}^3$ we use the following coordinate terminology:
\begin{align*}
{\mathbf{y}}=\sum_{i=1}^3 y_i {\mathbf{d_i}} =\sum_{i=1}^3 \breve y_i {\mathbf{a_i}}
\end{align*}
with $\mathsf{y}=(y_1, y_2, y_3)\in \mathbb{R}^3$ and $\breve{\mathsf{ y}}=(\breve{y}_1, \breve{y}_2, \breve{y}_3)\in \mathbb{R}^3$. Hence, the coordinate triples fulfill $\mathsf{y}=\mathsf{R}\cdot \breve{\mathsf{y}}$ with the associated orthogonal matrix $\mathsf{R}=({\mathbf{d_i}}\cdot {\mathbf{a_j}})_{ij}\in SO(3)$.

Proceeding from a three-dimensional free boundary problem (BVP) the derivation of a Cosserat rod model describing dry spinning of a single curved fiber can be done with the same methods and concepts as presented for an uni-axial fiber model in \cite{wieland:p:2019}. Nevertheless, the asymptotic considerations become more technical and lengthy. Here, we omit such a detailed derivation and formulate our dry spinning model straightforwardly.

Let $s$ and $r$ address the scaled arc length and radial parameters. Then, our one-two-dimensional dry spinning Cosserat rod model reads in non-dimensional form
\begin{system}[One-two-dimensional BVP for curved radially symmetric fibers]\label{sec:model_sys:rod}~\\
One-dimensional equations, $s\in (0,1)$:
\begin{equation}\label{sec:model_eq:1drod}
\begin{aligned}
L^{-1}\partial_s\breve{\mathsf{r}} &= \mathsf{R}^T\cdot\mathsf{e}_3,\\
L^{-1}\partial_s\mathsf{R} &= -\kappa\times\mathsf{R},\\
L^{-1}\partial_s \mathsf{n} &= \varrho_M u^2\kappa\times\mathsf{e}_3 + \mathsf{n}\times\kappa + L^{-1}\varrho_M u(\partial_s u)\mathsf{e}_3\\
&\quad - \mathsf{R}\cdot\breve{\mathsf{f}},\\
L^{-1}\partial_s \mathsf{m} &= L^{-1}\varepsilon^2 \frac{\varrho_M^2}{\rho}\mathsf{P_2}\cdot(u\partial_s(u \kappa)-\kappa u\partial_s u)\\
&\quad  + \varepsilon^2 \frac{\varrho_M^2}{\rho} u^2 \kappa\times\mathsf{P_2}\cdot\kappa + \mathsf{n}\times\mathsf{e}_3 + \mathsf{m}\times\kappa,\\
L^{-1}\partial_s u &= \mathrm{Re}\frac{1}{3\langle\mu(c,T)\rangle_{R^2}}n_3,\\
L^{-1}\partial_s(u\kappa) &= \frac{\mathrm{Re}}{\varepsilon^2}\frac{\rho}{3\varrho_M\langle\mu(c,T) \rangle_{R^2}}\mathsf{P_{2/3}}^{-1}\cdot\mathsf{m},
\end{aligned}
\end{equation}
with boundary conditions at inlet $s=0$ and fiber end $s=1$:
\begin{align*}
u(0) &= u_{in}, \quad  &u(1) &= \mathrm{Dr}, \quad &\breve{\mathsf{r}}(0) &= \breve{\mathsf{r}}_{in},\quad & \breve{\mathsf{r}}(1) &= \breve{\mathsf{r}}_{out},\\ \mathsf{R}(0) &= \mathsf{R}_{in}, \quad &\kappa(0) &= \kappa_{in},\quad &\kappa(1) &= \kappa_{out}.
\end{align*}
Radial equations, $(r,s)\in (0,1)^2$:
\begin{equation}\label{sec:model_eq:2drod}
\begin{aligned}
L^{-1}u \partial_s c - \frac{1}{\varepsilon\mathrm{Pe}_M}\frac{\bar{c}D(\bar{c},\bar{T})}{R^2 r}\partial_{r}(r\partial_r c) &= 0,\\
L^{-1}\rho(\bar{c},\bar{T}) q(\bar{c},\bar{T}) u \partial_s T  - \frac{1}{\varepsilon\mathrm{Pe}_T} \frac{C(\bar{c},\bar{T})}{R^2 r}\partial_r(r\partial_r T) &= 0,
\end{aligned}
\end{equation}
with boundary conditions at inlet $s=0$, fiber surface $r=1$ and symmetry boundary $r=0$:
\begin{align*}
c\big|_{s=0} &= c_{in}, \qquad \qquad \partial_r c\big|_{r=0} = 0,\\
\frac{1}{\mathrm{Pe}_M}\frac{\rho(\bar{c},\bar{T})D(\bar{c},\bar{T})}{R}\partial_r c\big|_{r=1} &= \mathrm{St}_M j_M(c,T)\big|_{r=1},\\
T\big|_{s=0} &= T_{in}, \qquad \qquad \partial_r T\big|_{r=1} = 0,\\
-\frac{1}{\mathrm{Pe}_T}\frac{C(\bar{c},\bar{T})}{R}\partial_{r} T\big|_{r=1} &= \big(\mathrm{St}_M j_M(c,T)(\delta(T)-h_d^0(T))\\
& \quad + \mathrm{St}_T j_T(T)\big)\big|_{r=1},\\
j_M(c,T) &= -\gamma(c,T)(c - c_{ref}(c,T)),\\
j_T(T) &= \alpha(T - T_\star).
\end{align*}
Constitutive laws and geometric relation:
\begin{equation*}
\begin{aligned}
\rho^{-1}(c,T) &= c\,(\rho_p^0)^{-1}(T) + (1-c)\,(\rho_d^0)^{-1}(T),\\
q(c,T) &= cq_p^0(T) + (1-c)q_d^0(T), \qquad q_d^0(T)=\partial_T h_d^0(T),\\
R(s) &= \sqrt{ \frac{Q}{\pi(\bar{c}\rho(\bar{c},\bar{T}) u)|_s}}.
\end{aligned}
\end{equation*}
Abbreviations:
\begin{align*}
\bar{c} &= \frac{1}{\pi}\langle c \rangle_{R^2},\quad \bar{T} = \frac{1}{\pi}\langle T \rangle_{R^2},\\
\langle y \rangle_{R^2(s)} &= 2\pi \int\limits_0^1 y(r,s)r\,dr, \quad \mathsf{P}_x = \frac{1}{4\pi}\mathrm{diag}(1,1,x).
\end{align*}
\end{system}
The rod equations \eqref{sec:model_eq:1drod} for the fiber curve $\mathsf{\breve{r}}$, triad $\mathsf{R}$, scalar speed $u$, curvature $\kappa$, contact force $\mathsf{n}$, and contact couple $\mathsf{m}$ describe the tangential fiber behavior parameterized by $s$. Here, $\varrho_M$ denotes the fiber mass line density, $R$ the radius, $\mathsf{\breve{f}}$ the body force line density, and $\mu$ the dynamic mixture viscosity. The assumption of an ideal mixture leads to the constitutive laws for the mixture density $\rho$ and the mixture specific heat capacity $q$. Here, $\rho_p^0$ and $\rho_d^0$ are the material densities and $h_p^0$ and $h_d^0$ the enthalpies of pure polymer and diluent, respectively. The temperature derivatives of $h_p^0$, $h_d^0$ and $h$ are in particular the specific heat capacities $q_p^0$, $q_d^0$, and $q$ for constant pressure. The two-dimensional advection-diffusion equations \eqref{sec:model_eq:2drod} for polymer mass fraction $c$ and temperature $T$ describe radial effects due to diluent evaporation (cf.~\cite{wieland:p:2019}) in dependence of $(r,s)$, where $D$ denotes the diffusion coefficient and $C$ the thermal conductivity. At the lateral fiber surface $j_M$, $j_T$ indicate the diluent mass and heat flux due to evaporation. At this surface the diluent density has a jump, which we formulate in terms of the mass fraction associated transfer coefficient $\gamma$ and the referential mass fraction in air $c_{ref}$ \cite{wieland:p:2019}. Whereas the temperature is continuous at the fiber surface, the heat flux has also a jump because of the heat exchange in the air due to the solvent evaporation with evaporation enthalpy $\delta$ of the diluent. The heat flux is described by the difference of the temperature at the fiber surface and away from the fiber $T_\star$ with heat transfer coefficient $\alpha$. Moreover, we introduce the constant polymer flux $Q=\bar{c}\varrho_M(\bar{c},\bar{T})u = c_{in}\varrho_{M,in}u_{in}$, where $\bar{c}$ and $\bar{T}$ denote the polymer mass fraction $c$ and temperature $T$ averaged over the fiber cross-sections. The subscript~$_{in}$ indicates the corresponding value at the nozzle. With $\mathsf{e_3}$ we denote the third canonical basis vector in $\mathbb{R}^3$. The reference values used for non-dimensionalization and the resulting dimensionless numbers are given in Tab.~\ref{sec:model_tab:refValues}. We fix $s_0 = L^\diamondsuit$ and $r_0 = \lVert \breve{\mathsf{r}}_{out}^\diamondsuit - \breve{\mathsf{r}}_{in}^\diamondsuit \rVert$, where the superscript $^\diamondsuit$ indicates dimensional quantities. Here, the label $^\diamondsuit$ is used for clarity, but suppressed in the following. The meaning of each quantity (dimensional or non-dimensional) will be clear from the context.

\begin{table}[t!]
\begin{minipage}[c]{0.48\textwidth}
\begin{center}
\begin{tabularx}{8.5cm}{| l r@{ = } l X |}
\hline
\multicolumn{4}{|l|}{\textbf{Reference values}}\\
Description & \multicolumn{2}{l}{Formula} & Unit\\
\hline
Length & $L_0$ & $r_0$ & m\rule{0pt}{2.6ex}\\
Radius & $R_0$ & $d_0$ & m\\
Scalar speed & $u_0$ & $v_0$ & m/s\\
Mass density & $\rho_0$ & $\varrho_{M0}/d_0^2$ & kg/m$^3$\\
Curvature & $\kappa_0$ & $1/r_0$ & 1/m\\
Stress & $n_0$ & $\varrho_{M0}v_0^2$ & N\\
Torque & $m_0$ & $\varrho_{M0}r_0v_0^2$ & N\,m\\
Outer force & $f_0$ & $\varrho_{M0}v_0^2/r_0$ & N/m\\
Enthalpy & $h_0$ & $q_0T_0$ & J/kg\\
Evaporation enthalpy & $\delta_0$ & $h_0$ & J/kg\\
Mass transfer coefficient & $\beta_0$ & $\gamma_0/\rho_{\star,0}$ & m/s\\
Air velocity & $v_{\star,0}$ & $v_0$ & m/s\\
Air temperature & $T_{\star,0}$ & $T_0$ & K\\
\hline
\end{tabularx}
\end{center}
\end{minipage}
\vfill
\vspace*{0.2cm}
\begin{minipage}[c]{0.48\textwidth}
\begin{center}
\begin{tabularx}{8.5cm}{| l r@{ = } X |}
\hline
\multicolumn{3}{|l|}{\textbf{Dimensionless numbers}}\\
Description & \multicolumn{2}{l|}{Formula}\\
\hline
Slenderness & $\varepsilon$ & $d_0/r_0$ \rule{0pt}{2.6ex}\\
Reynolds & $\mathrm{Re}$ & $\varrho_{M0}v_0r_0/(d_0^2\mu_0)$\\
Froude & $\mathrm{Fr}$ & $v_0/\sqrt{gr_0}$ \\
Mass Peclet & $\mathrm{Pe}_M$ & $v_0d_0/D_0$\\
Temperature Peclet & $\mathrm{Pe}_T$ & $\varrho_{M0}v_0q_0/(C_0d_0)$\\
Mass Stanton & $\mathrm{St}_M$ & $\gamma_0d_0^2/(v_0\varrho_{M0})$\\
Temperature Stanton & $\mathrm{St}_T$ & $\alpha_0d_0^2/(v_0\varrho_{M0}q_0)$\\
Drawing & $\mathrm{Dr}$ & $u_{out}/u_0$\\
Air drag associated & $\mathrm{A}_\star$ & $\rho_{\star,0}d_0v_0^2/f_0$\\
Air-fiber Reynolds & $\mathrm{Re}_\star$ & $d_0v_0/\nu_{\star,0}$\\
Nusselt & $\mathrm{Nu}_\star$ & $\alpha_0d_0/\lambda_{\star,0}$\\
Prandtl & $\mathrm{Pr}_\star$ & $q_{\star,0}\rho_{\star,0}\nu_{\star,0}/\lambda_{\star,0}$\\
Sherwood & $\mathrm{Sh}_\star$ & $\gamma_0d_0/(\rho_{\star,0}D_{d,\star,0})$\\
Schmidt & $\mathrm{Sc}_\star$ & $\nu_{\star,0}/D_{d,\star,0}$\\
\hline
\end{tabularx}
\end{center}
\end{minipage}
\vspace*{0.2cm}
\caption{Composite reference values used for non-dimensionalization and resulting dimensionless numbers. The following scales are assumed to be given from the specific considered setup $\varrho_{M0}$, $v_0$, $r_0$, $d_0$, $\mu_0$, $q_0$, $T_0$, $\alpha_0$, $\gamma_0$, $C_0$, $D_0$, $\rho_{\star,0}$, $\nu_{\star,0}$, $p_{\star,0}$ $q_{\star,0}$, $\lambda_{\star,0}$, $D_{d,\star,0}$.}\label{sec:model_tab:refValues}
\end{table}

As external forces we consider gravitational and air drag forces $\breve{\mathsf{f}} = \breve{\mathsf{f}}_g + \breve{\mathsf{f}}_{air}$
\begin{align*}
\breve{\mathsf{f}}_g &= \frac{1}{\mathrm{Fr}^2}\varrho_M\breve{\mathsf{e}}_g,\\
\breve{\mathsf{f}}_{air} &= \frac{\mathrm{A}_\star}{\mathrm{Re}_\star^2}\frac{\rho_\star\nu^2_\star}{2R} \breve{\mathsf{F}}\bigg(\breve{\mathsf{t}},\mathrm{Re}_\star\frac{2R}{\nu_\star}\breve{\mathsf{v}}_{rel}\bigg),
\end{align*}
with direction of gravity $\breve{\mathsf{e}}_g$, $\lVert\breve{\mathsf{e}}_g\rVert = 1$, and dimensionless drag function $\breve{\mathsf{F}}$ given in \cite{marheineke:p:2009b}. The normalized fiber tangent and the relative velocity read in outer basis $\breve{\mathsf{t}} = \mathsf{R}^T\cdot\mathsf{e_3}$ and $\breve{\mathsf{v}}_{rel} = \breve{\mathsf{v}}_\star-\breve{\mathsf{v}}$ with fiber velocity $\breve{\mathsf{v}} = u\mathsf{R}^T\cdot\mathsf{e_3}$. Note that to distinguish the fiber quantities from the airflow quantities all airflow associated fields are labeled with the index $_\star$. In particular, $\breve{\mathsf{v}}_\star$ denotes the velocity, $\rho_\star$ the density, $\nu_\star$ the kinematic viscosity, $\lambda_\star$ the thermal conductivity, and $q_\star$ the specific heat capacity of the air. Moreover, $D_{d,\star}$ denotes the diffusivity of diluent in the air and $\rho_{d,\star}$ the diluent density in the air away from the fiber. All these quantities are space- and time-dependent fields assumed to be dimensionless and known -- for example provided by an external computation. The corresponding reference values used for non-dimensionalization are denoted with the index $_0$ and given in Table~\ref{sec:model_tab:refValues}. Furthermore, we employ the models for the heat and mass transfer
\begin{align*}
\alpha &= \frac{1}{\mathrm{Nu}_\star}\frac{\lambda_\star}{2R}\mathcal{N}\left(\mathrm{Re}_\star\frac{2R}{\nu_\star}\breve{\mathsf{v}}_{rel}\cdot\breve{\mathsf{t}}, \mathrm{Re}_\star\frac{2R}{\nu_\star}\lVert \breve{\mathsf{v}}_{rel} \rVert, \mathrm{Pr}_\star\frac{q_\star\rho_\star\nu_\star}{\lambda_\star}\right),\\
\gamma &= \beta \varrho, \qquad c_{ref} = 1-\frac{\rho_{d,\star}}{\varrho},\\
\beta &= \frac{1}{\mathrm{Sh}_\star}\frac{D_{d,\star}}{2R}\mathcal{N}\left(\mathrm{Re}_\star\frac{2R}{\nu_\star}\breve{\mathsf{v}}_{rel}\cdot\breve{\mathsf{t}}, \mathrm{Re}_\star\frac{2R}{\nu_\star}\lVert \breve{\mathsf{v}}_{rel} \rVert, \mathrm{Sc}_\star\frac{\nu_\star}{D_{d,\star}}\right)
\end{align*}
with the associated dimensionless function $\mathcal{N}:\mathbb{R}^3\rightarrow\mathbb{R}$ and the model for the diluent density in the air at the fiber surface $\varrho$ given in \cite{wieland:p:2019}. For an appropriate closing of the System~\ref{sec:model_sys:rod} we need rheological laws for the viscosity $\mu$, the evaporation enthaply $\delta$, the conductivity $C$, the diffusivity $D$, as well as the material densities $\rho_p^0$, $\rho_d^0$ and the material specific heat capacities $q_p^0$, $q_d^0$, which we will specify in the industrial setup.

Note that for the determination of the a priori unknown fiber length $L$ we impose an additional boundary condition: to the $17$ equations of \eqref{sec:model_sys:rod} for the $17$ unknowns $(\breve{\mathsf{r}}, \mathsf{R},  u, \kappa, \mathsf{n}, \mathsf{m})$, we set $18$ boundary conditions at $s=0$ and $s=1$.
The one-dimensional system \eqref{sec:model_eq:1drod} is the main difference between the curved fiber model (System~\ref{sec:model_sys:rod}) and its uni-axial counterpart given in \cite{wieland:p:2019}. Whereas for curved fibers it contains $18$ unknowns, the corresponding one-dimensional uni-axial model is formulated with respect to only three unknowns $(u,n_3,L)$.

\section{Numerical scheme}\label{sec:numerics}
Besides the unknown parameter $L$, the curved fiber model (System~\ref{sec:model_sys:rod}) has the same structure as the uni-axial one in \cite{wieland:p:2019}, such that the numerical procedure can be adopted from that case. For given polymer mass fraction $c$ and mixture temperature $T$, the one-dimensional equations \eqref{sec:model_eq:1drod} form together with the boundary conditions a parametric boundary value problem of ordinary differential equations. For its solution we employ a continuation-collocation method similar to the one in \cite{wieland:p:2019}. Due to the higher complexity of the one-dimensional rod model \eqref{sec:model_eq:1drod} compared to its uni-axial string counterpart a suitable adaption of the model specific continuation procedure is required, which we discuss in the following. The solution of the one-dimensional equations is iteratively coupled with the solution of the two-dimensional equations \eqref{sec:model_eq:2drod}. A solution of the two-dimensional equations \eqref{sec:model_eq:2drod} itself can implicitly be given in terms of Green's functions and Volterra integral equations of second kind with singular kernel for the values at the fiber boundary. To achieve feasible computation times in view of the further two-way coupling with an airflow simulation, we employ the same product integration method based on the Lobatto~IIIa quadrature formula as in the uni-axial case. The fiber-air interaction is performed by a weakly coupling of the fiber calculation and airflow computation via iterative solving. This means that our numerical solution procedure consists of nested iterations: an inner iteration for the fiber solution (coupling of one and two-dimensional fiber equations) and an outer iteration for the coupling of fiber and airflow solutions.

In this paper we describe the new model-dependent continuation strategy for the solution of the one-dimensional equations \eqref{sec:model_eq:1drod}. For a detailed description of the product integration method for the solution of the two-dimensional equations \eqref{sec:model_eq:2drod} and the implementation of the fiber-airflow interactions we refer to \cite{wieland:p:2019}.

A direct solution of the boundary value problem \eqref{sec:model_eq:1drod} with the help of classical solvers such as collocation schemes is in general not possible due to the non-global convergence of the Newton method. Therefore, the system \eqref{sec:model_eq:1drod} is embedded into the family of boundary value problems
\begin{align*}
\frac{\mathrm{d}}{\mathrm{d}s}\mathbf{y}=\hat{\mathbf{f}}(\mathbf{y};\mathbf{p}),  \qquad & \hat{\mathbf{g}}(\mathbf{y}(0),\mathbf{y}(1);\mathbf{p})=\mathbf{0},\\
\hat{\mathbf{f}}(\cdot;\mathbf{\underline 1})=\mathbf{f}, \qquad \hat{\mathbf{g}}(\cdot,\cdot;\mathbf{\underline 1})=\mathbf{g}, \qquad & \hat{\mathbf{f}}(\cdot;\mathbf{0})=\mathbf{f}_0, \qquad \hat{\mathbf{g}}(\cdot,\cdot;\mathbf{0})=\mathbf{g}_0,
\end{align*}
with continuation tuple $\mathbf{p}\in[0,1]^n$. Here, $\mathbf{\underline 1}$ denotes the $n$-dimensional tuple of ones. The functions $\mathbf{f}_0$, $\mathbf{g}_0$ are chosen in such a way that for $\mathbf{p}=\mathbf{0}$ an analytical solution is known. Given this starting solution, we seek for a sequence of parameter tuples $\mathbf{0}=\mathbf{p}_0, \mathbf{p}_1, \ldots, \mathbf{p}_m = \mathbf{\underline 1}$ such that the solution to the respective predecessor boundary value problem provides a good initial guess for the successor. The solution associated to $\mathbf{p}= \mathbf{\underline 1}$ finally belongs to the original system. Each boundary value problem itself is solved by a collocation scheme based on the Lobatto~IIIa formula. This is a classical approach provided in the software MATLAB (\texttt{http://www.mathworks.com}) by the routine \texttt{bvp4c.m}.

We introduce continuation parameters for the viscous, gravitational and aerodynamic forces as well as for the draw ratio and the effect of heat and mass transfer over the fiber boundary. Additionally, we introduce the further continuation parameter $p_{bc}$ and the modified boundary condition
\begin{align*}
\breve{\mathsf{r}}(0) - p_{bc}\breve{\mathsf{r}}_{in} - (1-p_{bc})\left(\breve{\mathsf{r}}_{out} - L\mathsf{R}_{in}^T\cdot\mathsf{e}_3\right)= 0.
\end{align*}
We choose the starting solution (corresponding to $\mathbf{p} = \mathbf{0}$)
\begin{equation*}
\begin{aligned}
\breve{\mathsf{r}}(s) &= \breve{\mathsf{r}}_{out}- (1-s)L\mathsf{R}_{in}^T\cdot\mathsf{e}_3, \qquad &\mathsf{R} &= \mathsf{R}_{in}, \\
\mathsf{n} &= 0,\qquad \mathsf{m} = 0,\qquad  u = u_{in},\\
\kappa &= 0,\qquad \bar{c} = c_{in}, \qquad \bar{T}=T_{in},
\end{aligned}
\end{equation*}
with $L  = (\breve{\mathsf{r}}_{out} - \breve{\mathsf{r}}_{in})\cdot(\mathsf{R}_{in}^T\cdot\mathsf{e}_3)$ and $s\in[0,1]$. Basically the initial solution for the rod curve $\breve{\mathsf{r}}$ 
is built by the orthogonal projection of $(\breve{\mathsf{r}}_{out} - \breve{\mathsf{r}}_{in})$ onto the initial fiber direction $\mathsf{R}_{in}^T\cdot\mathsf{e}_3$. In total, the initial solution is a stress-free, unbent, straight fiber with constant polymer mass fraction, speed and temperature being extruded from the modified nozzle position $(\breve{\mathsf{r}}_{out} - L\mathsf{R}_{in}^T\cdot\mathsf{e}_3)$. The navigation strategy through the parameter space equals the corresponding procedure for the uni-axial fiber model.

\begin{remark}[Regularization of rod model]
Numerical studies show that the angular momentum equation of the rod model \eqref{sec:model_eq:1drod} is badly scaled with respect to the variable $\mathsf{m}$. In order to accelerate the convergence of the continuation-collocation procedure we introduce the scaled couple $\tilde{\mathsf{m}} = \mathsf{m}/\varepsilon^2$, yielding the angular momentum balance
\begin{align*}
L^{-1}\partial_s \tilde{\mathsf{m}} =& \frac{\varrho_M^2}{\rho} u^2 \kappa\times\mathsf{P_2}\cdot\kappa + L^{-1} \frac{\sigma_M^2}{\rho}\mathsf{P_2}\cdot(-\kappa u\partial_s u + u\partial_s(u \kappa))\\
& + \frac{1}{\varepsilon^2}\mathsf{n}\times\mathsf{e}_3 + \tilde{\mathsf{m}}\times\kappa.
\end{align*}
Here, the slenderness parameter $\varepsilon$ only appears in the term $1/\varepsilon^2\mathsf{n}\times\mathsf{e}_3$. It is considered as regularization parameter (regularization of the associated string model) and therefore fixed at the constant value $\varepsilon_0 = 10^{-1}$. The remaining parameters $\varepsilon$ in the two-dimensional equations \eqref{sec:model_eq:2drod} are not affected and set as desired. The corresponding material law in \eqref{sec:model_eq:1drod} changes accordingly.
\end{remark}

\section{Dry spinning of curved fibers with two-way coupled airflow}\label{sec:ex}
As in \cite{wieland:p:2019} we consider a cellulose acetate (CA)-acetone mixture being vertically dry spun in a spinning duct. In difference the air inflow streams horizontally through the spinning chamber instead of vertically, which is a typical situation in industrial setups. Thus, the fibers are exposed to a perpendicular air inflow situation. This leads to lateral movements of the fibers such that an uni-axial fiber model is not sufficient for this flow situation. Hence, we employ our curved fiber model (System~\ref{sec:model_sys:rod}). In the following we describe the process setup and present simulation results for the curved fibers with two-way coupled airflow.

\subsection{Process setup}
\begin{figure}[!t]
\begin{center}
\includegraphics[height=8cm]{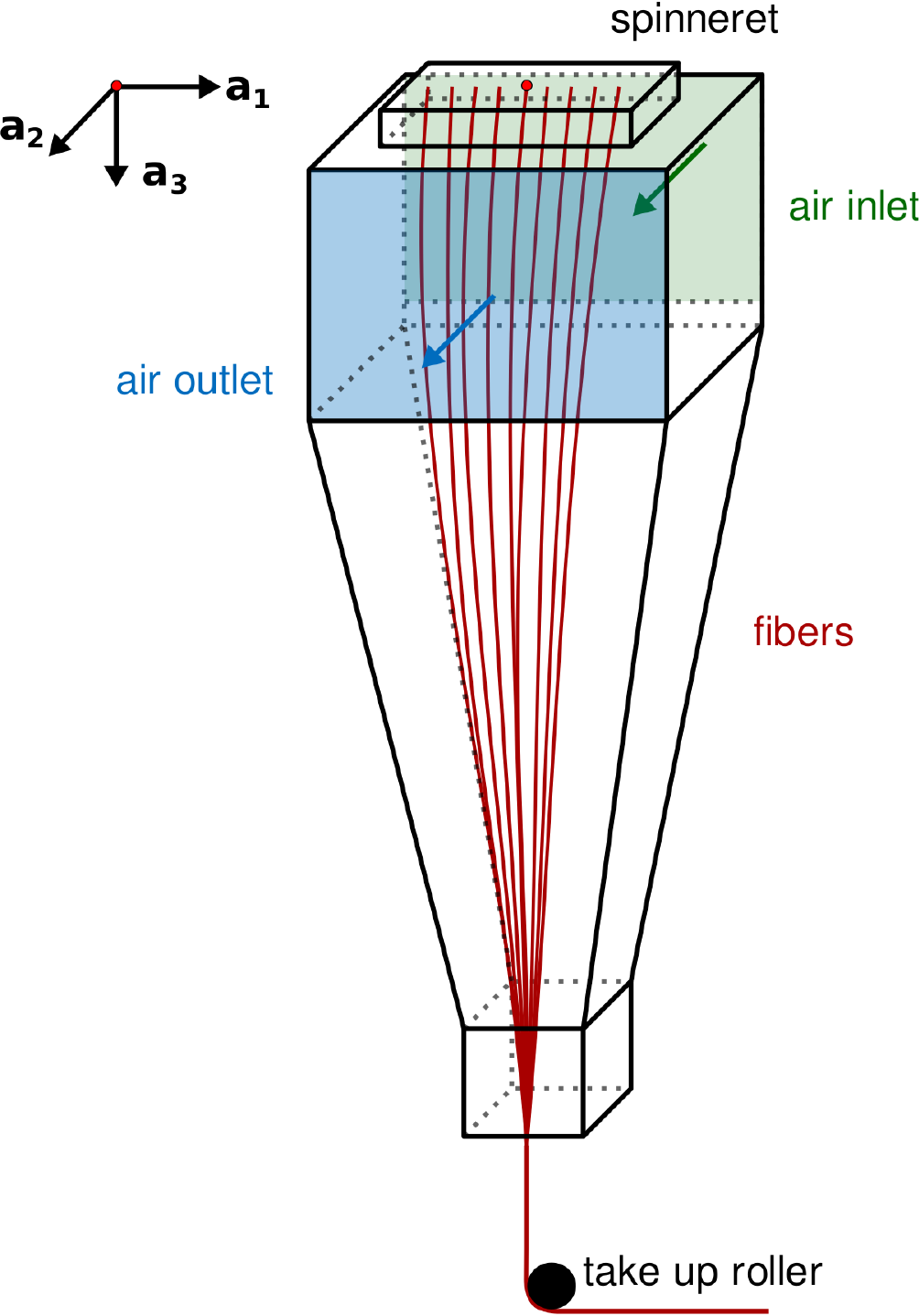}
\hspace*{0.5cm}
\includegraphics[height=2.25cm]{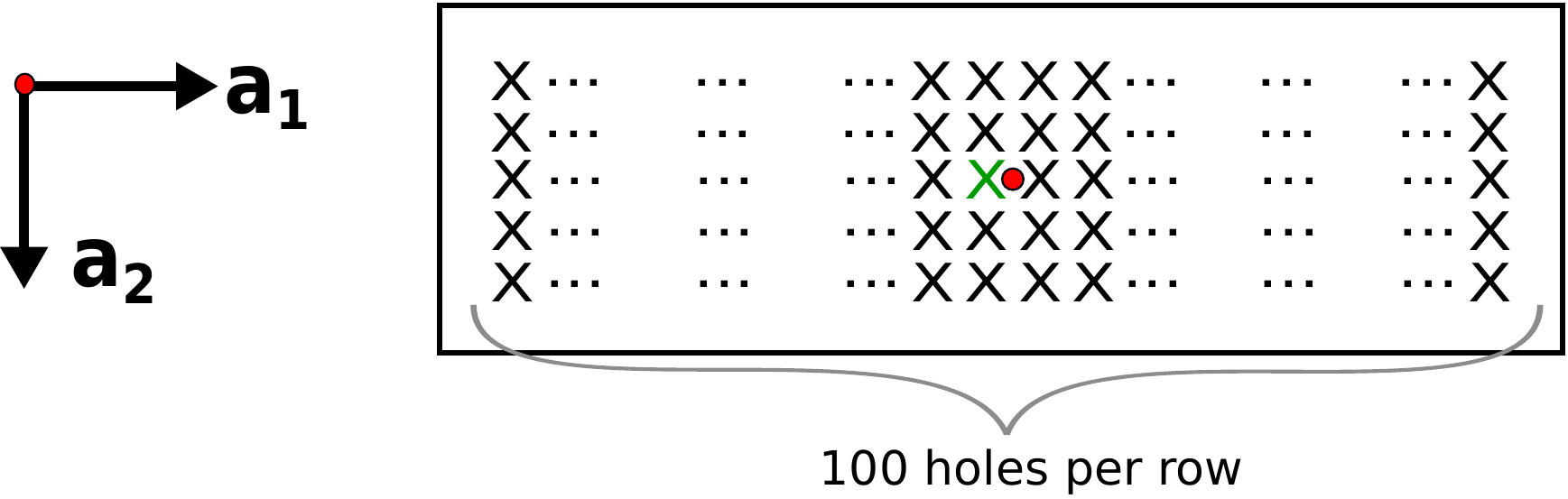}
\end{center}
\caption{\textit{Left:} Considered dry spinning device with horizontal air inflow (in $\mathbf{a_2}$-direction) and one row of spun fibers. \textit{Right:} Top view of the holes in the spinneret \textit{(marked with $\mathsf{X}$)} (not in scale). All proceeding simulation results are shown for the fiber extruded from the central position \textit{(green $\mathsf{X}$)}. The origin of the coordinate axis is marked with a \textit{red point}. }\label{sec:ex_fig:apparatus_curved}
\end{figure}
We consider a spinning duct with the air inflow situation as depicted in Fig.~\ref{sec:ex_fig:apparatus_curved}. The device geometry is adopted from an actual industrial dry spinning device. To show the capability of our numerical framework we let 500 fibers being spun simultaneously from the holes of a spinneret into the spinning chamber. These holes are ordered in five parallel rows with 100 holes each, as sketched in Fig.~\ref{sec:ex_fig:apparatus_curved}. We consider the fibers to be injected from the plane $\{\mathsf{\breve{x}}\in\mathbb{R}^3~|~ \breve{x}_3 = 0\}$ and the middle row additionally to lie in $\breve{x}_2 = 0$. At the bottom of the device all fibers are taken up by a roller. Concerning the airflow we assume dry air at the inlet, i.e., the inflow contains no diluent. The absolute air inflow velocity (in $\mathbf{a_2}$-direction) at the air inlet is $\|\mathbf{v}_\star\| = 0.22$~m/s and the air inflow temperature is $T_\star=330$~K. The remaining process parameters are given in Table~\ref{sec:ex_table:param}.
\begin{table}[!t]
\begin{minipage}[t]{0.49\textwidth}
\begin{center}
\begin{tabularx}{8cm}{| l r@{ = } X |}
\hline
\multicolumn{3}{|l|}{\textbf{Process parameters}}\\
Description & \multicolumn{2}{l|}{Value}\\
\hline
Device height & $H$ & $5.35$ m\rule{0pt}{2.6ex}\\
Nozzle radius & $R_{in}$ & $1\cdot10^{-4}$ m\\
Speed at nozzle & $u_{in}$ &$5$ m/s\\
Temperature at nozzle & $T_{in}$ & $ 350$ K\\
Polym. mass fract. at nozzle & $c_{in}$ & $0.29$\\
Take up speed at bottom & $u_{out}$& $10$ m/s\\
Total number of fibers & $M$& $500$\\
\hline
\end{tabularx}
\end{center}
\end{minipage}
\vfill\vspace*{0.2cm}
\begin{minipage}[t]{0.49\textwidth}
\begin{center}
\begin{tabularx}{8cm}{| l r@{ = } X |}
\hline
\multicolumn{3}{|l|}{\textbf{Physical parameters}}\\
Description & \multicolumn{2}{l|}{Value}\\
\hline
CA density & $\rho_p^0$ &$1300$ kg/m$^3$\rule{0pt}{2.6ex}\\
Acetone density & $\rho_d^0 $&$ 767$ kg/m$^3$\\
CA spec. heat cap. & $q_{p}^0$ &$1600$ J/(kg\,K)\\
Acetone spec. heat cap. & $q_{d}^0 $&$2160$ J/(kg\,K)\\
\hline
\end{tabularx}
\end{center}
\end{minipage}
\vfill\vspace*{0.2cm}
\caption{Process and physical parameters of the industrial dry spinning setup with horizontal air inflow (cf. Fig.~\ref{sec:ex_fig:apparatus_curved}).}\label{sec:ex_table:param}
\end{table}

The physical material parameters for the polymer CA and the diluent acetone are listed in Table~\ref{sec:ex_table:param}. Concerning the closing models for the viscosity $\mu$, the evaporation enthalpy $\delta$, the conductivity $C$, and the diffusivity $D$ we refer to \cite{wieland:p:2019}. The setup specific reference values used for the non-dimensional form of the model equations as well as the resulting dimensionless numbers are given in Tab.~\ref{sec:ex_table:dimNum}.
\begin{table}[!t]
\begin{minipage}[t]{0.48\textwidth}
\begin{center}
\begin{tabularx}{7cm}{| l r@{ = } X |}
\hline
\multicolumn{3}{|l|}{\textbf{Reference values}}\\
Description & \multicolumn{2}{l|}{Formula}\\
\hline
Mass line density & $\varrho_{M0}$ & $\rho(c_{in})R_{in}^2\pi$ \rule{0pt}{2.6ex}\\
Radius & $d_0$ & $2R_{in}$\\
Viscosity & $\mu_0$ & $\mu(c_{in},T_{in})$\\
Specific heat capacity & $q_0$ & $q(c_{in})$\\
Thermal conductivity & $C_0$ & $C(c_{in},T_{in})$\\
Diffusivity & $D_0$ & $D(c_{in},T_{in})$\\
Further scales & $b_{0}$ & $b_{in}$, \\
\multicolumn{3}{|l|}{\qquad $b \in \{v,T,\alpha,\gamma,p_\star,\lambda_\star,\nu_\star,q_\star,\rho_\star,D_{d,\star}\}$}\\
\hline
\end{tabularx}
\end{center}
\end{minipage}
\vfill\vspace*{0.2cm}
\begin{minipage}[!ht]{0.48\textwidth}
\begin{center}
\begin{tabularx}{7cm}{| l r@{ = } X |}
\hline
\multicolumn{3}{|l|}{\textbf{Dimensionless numbers}}\\
Description & \multicolumn{2}{l|}{Value}\\
\hline
Slenderness & $\varepsilon$ & $3.74\cdot 10^{-5}$ \rule{0pt}{2.6ex}\\
Reynolds & $\mathrm{Re}$ & $2.08\cdot10^3$\\
Froude & $\mathrm{Fr}$ & $6.90\cdot10^{-1}$ \\
Mass Peclet & $\mathrm{Pe}_M$ & $1.31\cdot 10^7$\\
Temperature Peclet & $\mathrm{Pe}_T$ & $8.69\cdot 10^3$\\
Mass Stanton & $\mathrm{St}_M$ & $2.25\cdot 10^{-4}$\\
Temperature Stanton & $\mathrm{St}_T$ & $3.11\cdot 10^{-5}$\\
Drawing & $\mathrm{Dr}$ & $2.00$\\
Air drag associated & $\mathrm{A_\star}$ & $3.77\cdot10^1$\\
Air-fiber Reynolds & $\mathrm{Re_\star}$ & $6.85\cdot10^1$\\
Nusselt & $\mathrm{Nu_\star}$ & $2.23$\\
Prandtl & $\mathrm{Pr_\star}$ & $7.44\cdot10^{-1}$\\
Sherwood & $\mathrm{Sh_\star}$ & $1.13\cdot10^1$\\
Schmidt & $\mathrm{Sc_\star}$ & $1.04$\\
\hline
\end{tabularx}
\end{center}
\end{minipage}
\vfill\vspace*{0.2cm}
\caption{Reference values and resulting dimensionless numbers for one representative fiber (cf. Fig.~\ref{sec:ex_fig:apparatus_curved}).}\label{sec:ex_table:dimNum}
\end{table}

\begin{figure}[!t]
\begin{minipage}[c]{0.95\textwidth}
	\centering
	\includegraphics{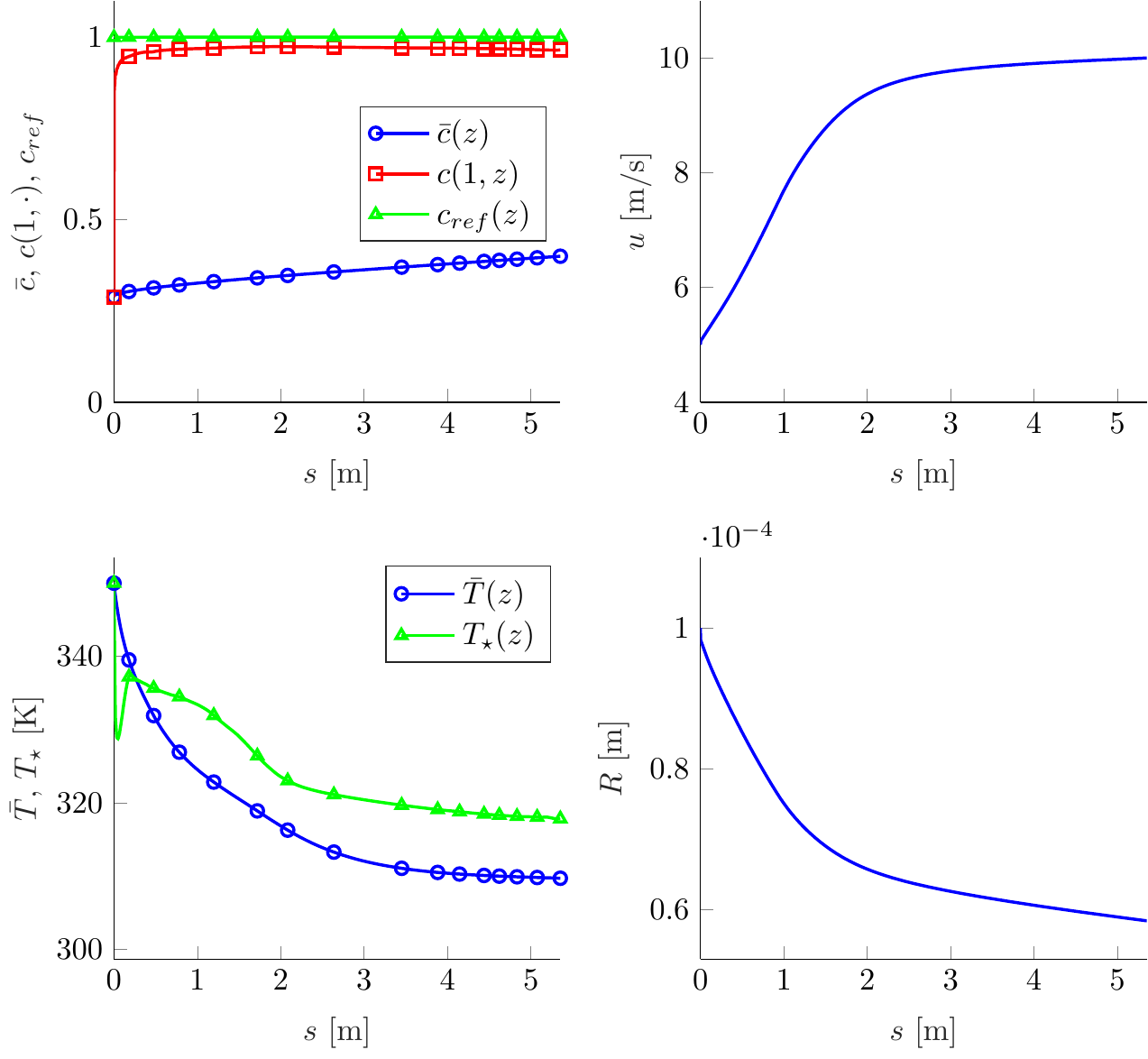}
\end{minipage}
\vfill\vspace*{-0.2cm}
\caption{Solution quantities of a centrally located fiber (cf.~Fig.~\ref{sec:ex_fig:apparatus_curved}). \textit{Top-left:} polymer (CA) mass fraction (averaged $\bar{c}$, at the fiber boundary $c\lvert_{r=1}$ and referential $c_{ref}$). \textit{Top-right:} scalar fiber speed $u$. \textit{Bottom-left:} averaged fiber temperature $\bar{T}$ and air temperature $T_\star$. \textit{Bottom-right:} resulting fiber radius $R$.}\label{sec:ex_fig:sol_c_u_T_R}
\end{figure}
\begin{figure}[!h]
\begin{minipage}[c]{0.95\textwidth}
	\centering
	\includegraphics{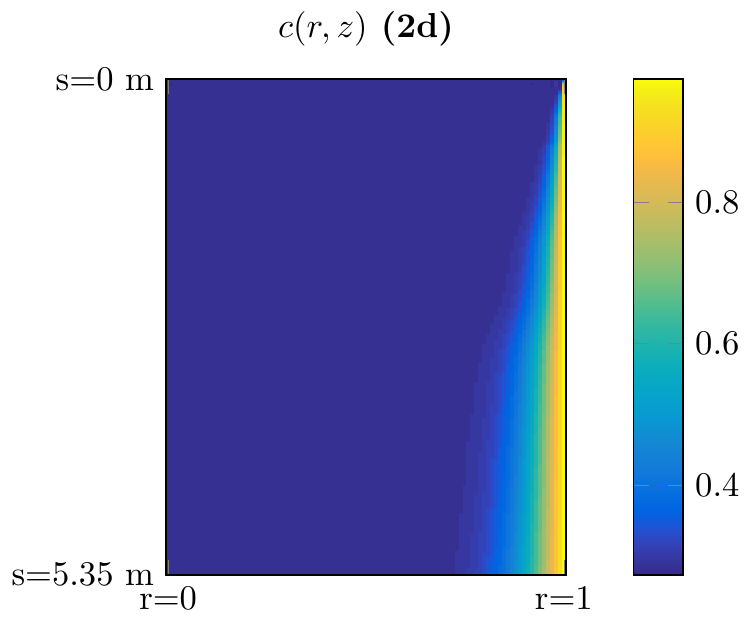}
\end{minipage}
\vspace*{-0.3cm}
\caption{Polymer (CA) mass fraction of fiber solution (cf.~Figs.~\ref{sec:ex_fig:apparatus_curved} and \ref{sec:ex_fig:sol_c_u_T_R}).}\label{sec:ex_fig:c_profile}
\end{figure}
\begin{figure}[!htbp]
\begin{minipage}[c]{0.95\textwidth}
	\centering
	\includegraphics{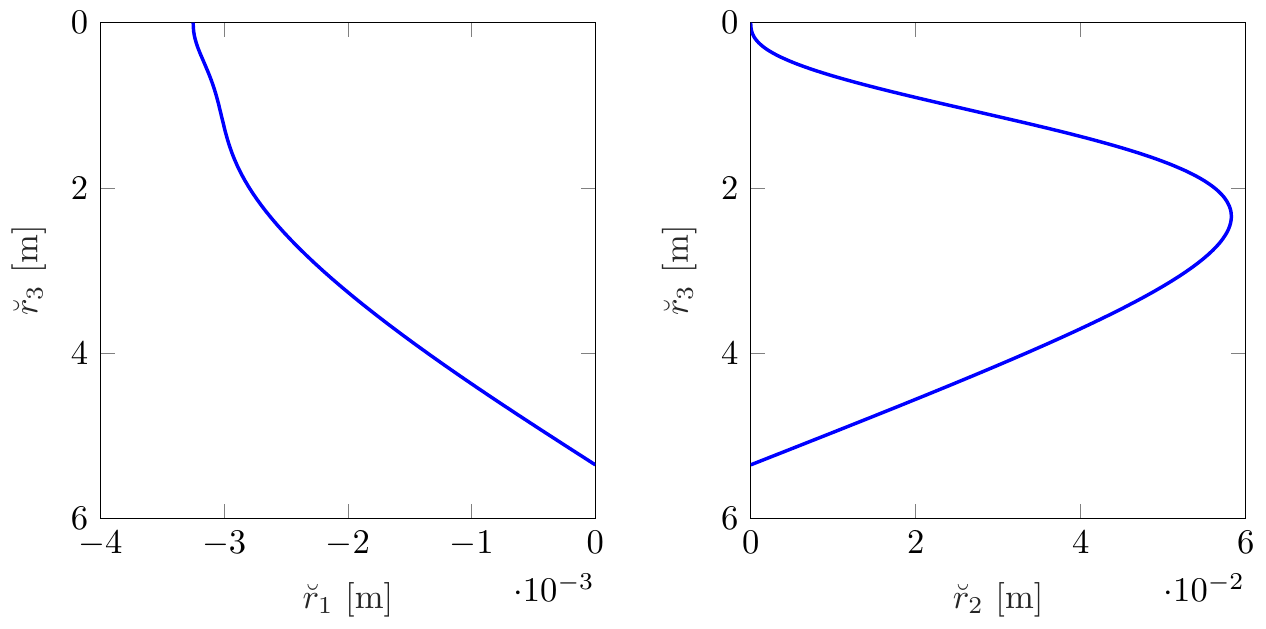}
\end{minipage}
\vspace*{-0.3cm}
\caption{Side views of the fiber curve of a centrally located fiber (cf.~Fig.~\ref{sec:ex_fig:apparatus_curved}).}\label{sec:ex_fig:sol_curve}
\end{figure}
\subsection{Results}
The forthcoming numerical simulations are performed on an Intel Core i7-6700 CPU
(4 cores, 8 threads) and 16 GBytes of RAM. The solver for the airflow meets an accuracy of $\mathcal{O}(10^{-6})$ and the break-up criterion of the fiber-air coupling algorithm satisfies an error tolerance $tol=10^{-5}$. After 10 iteration steps the fiber-air coupling algorithm is converged. The overall computation time for the presented setup is around 50 hours.

Since the solutions of the 500 fibers do not show visible differences, we exemplarily illustrate the solution behavior for the fiber that is centrally located in the spinning chamber (cf.\ Fig.~\ref{sec:ex_fig:apparatus_curved}). In Fig.~\ref{sec:ex_fig:sol_c_u_T_R} we see that
the cross-sectionally averaged CA mass fraction increases from $c_0 = 0.29$ at the inlet to $\bar{c} = 0.40$ at the outlet, indicating the evaporation of the acetone during the spinning process. The scalar fiber speed $u$ increases monotonically and reaches the take up speed ($u_{out}=10$~m/s) at the device end. Directly at the nozzle the averaged fiber temperature $\bar{T}$ starts to decrease due to the immediate set in of the acetone mass transfer caused by evaporation, which is indicated by the rapid rise of the CA mass fraction at the fiber boundary $c|_{r=1}$. In the nozzle region the airflow is heated due to the positive relative temperature between fiber and airflow, i.e., $(\bar{T}-T_\star) > 0$. Away from the nozzle the fibers cool the air due to a negative relative temperature, i.e., $(\bar{T}-T_\star) < 0$. As a result of the solvent evaporation the fiber reaches the minimal temperature $T = 310$~K which is less than the inflow air temperature $T_\star = 330$~K. The fiber radius decreases from its initial value $R_{in} = 10^{-4}$~m to $R=5.8\cdot10^{-5}$~m at the fiber end. This fiber thinning is caused by the fiber take up at the bottom of the spinning device as well as by the acetone evaporation.

Worth to investigate is the profile for the polymer mass fraction in Fig.~\ref{sec:ex_fig:c_profile} that indicates an inhomogeneous CA-acetone distribution in radial direction. While the fiber surface is nearly completely dried shortly away from the nozzle, the innermost part of the fiber contains the initial proportion of diluent over its complete length. In contrast there are no visible radial effects in the temperature and hence we omit the visualization of the fiber temperature at the boundary and the radial temperature profile.

Fig.~\ref{sec:ex_fig:sol_curve} visualizes the curve for the central fiber. Whereas the deflection in $\mathbf{a_1}$-direction is mainly due to the take up of the fiber at the central position at the bottom ($\breve{x}_1 = \breve{x}_2 = 0$), the deflection in $\mathbf{a_2}$-direction is due to the airflow, which causes tangential forces on the fiber in the upper part of the spinning duct leading to its lateral movement. The maximal deflection of the fiber in $\mathbf{a_2}$-direction is $5.8\cdot10^{-2}$~m.

\section{Conclusion}
In this work we develop a dry spinning model for curved viscous fibers. The model combines one-dimensional rod equations with two-dimensional advection-diffusion equations covering the radial effects of mass fraction and temperature. Adopting the efficient numerical framework from the uni-axial case \cite{wieland:p:2019} allows simulations of industrial dry spinning setups with two-way coupled fiber-air interactions, where lateral fiber movements take place. The efficiency of the proposed model-simulation framework builds a good basis for optimization and optimal design of dry spinning processes.

%

\end{document}